\documentclass[useAMS,usenatbib]{mn2e}

\usepackage{amsmath}
\usepackage{mathtools}
\usepackage{amssymb,amsfonts,amssymb}  
\usepackage{latexsym}                           
\usepackage[letterpaper]{geometry} 
\usepackage{graphicx}
\title[Binary stars in population synthesis models]{Revisiting binary stars in population synthesis models}
\author[F. Hern\'andez-P\'erez and G. Bruzual]{Fabiola Hern\'andez-P\'erez$^{1,2}$\thanks{E-mail:
fhernandez@cida.ve} and Gustavo Bruzual A.$^{3}$\\
$^{1}$Centro de Investigaciones de Astronom\'ia, CIDA, Av. Alberto Carnevalli,  M\'erida, Venezuela. A.P. 264, C.P. 5101\\
$^{2}$Postgrado de F\'isica Fundamental, Universidad de Los Andes, M\'erida, Venezuela\\
$^{3}$Centro de Radioastronom\'ia y Astrof\'isica, CRyA, UNAM, Campus Morelia, Michoac\'an. M\'exico. A.P. 3-72, C.P.58089}
\begin{document}

\date{Accepted 2013 February 22.  Received 2013 February 22; in original form 2012 December 20}

\pagerange{\pageref{firstpage}--\pageref{lastpage}} \pubyear{2013}

\maketitle
\label{firstpage}

\begin{abstract}
We report results of a population synthesis model that follows the evolution of single and binary stars.
In this model we include the 2HeWD merger channel, suggested by \cite{han02},  for the formation of EHB stars.
The physical parameters of the resulting EHB stars are derived from the BaSTI database, and are thus
realistic and observationally supported.
The predictions of this model are in good agreement with traditional population synthesis models, except
when the spectrum of the stellar population is dominated by binary stars or their products, e.g., EHB stars
in the UV of ETGs. We reproduce successfully the observed CMD and SED of the metal rich open cluster NGC 6791.
The stellar population in this cluster may be archetypal of the stellar population in ETGs that show the
UVX phenomenon. Our models should be appropriate to study the UV upturn in ETGs.
\end{abstract}


\section{Introduction}
{\label{sec:intro}}

Stellar population synthesis models are a useful tool to interpret the spectrum and colours of the light emitted by galaxies, 
as well as the properties of resolved stellar populations in the colour-magnitude diagram (CMD).
Crucial ingredients for these models are complete sets of evolutionary tracks that describe in detail the time dependence
of the physical properties of stars of different initial mass and chemical composition in the Hertzsprung-Russell diagram (HRD).
Equally complete spectral libraries are needed to compute the spectrophotometric properties of the stars at each position in the HRD.
Weighting the stellar spectrum by the number of stars at each of these positions at any given time, we obtain the integrated spectrum 
of the stellar population at this age.
The number of stars of each mass born whenever there is a star formation event is given by the \textit{initial mass function} (IMF),
and the mass of gas transformed into stars as a function of time in each of these events is provided by the \textit{star formation rate} (SFR). 
See, e.g., \cite{mf97}, \cite{bc03}, hereafter BC03, and \cite{cm05} for details.

Despite considerable progress in recent years in both the quality and quantity of the ingredients
available to the population synthesis modelers, i.e., stellar tracks and spectra, 
these models still suffer from limitations in some specific regimes.
In general, the problems arise from our poor understanding of short-lived and/or not well characterized phases of stellar evolution,
e.g., the thermally pulsing asymptotic giant branch (TP-AGB) and the extreme horizontal branch (EHB).
These stars are relatively bright and contribute considerably to the total luminosity of a simple stellar population (SSP).
Thanks to the work of P. Marigo and collaborators \citep{pm08}, we now have a better understanding
of TP-AGB stars of intermediate mass.
These stars contribute at least 50\% of the near infrared (NIR) light in a 1-2 Gyr old SSP  \citep{gb13}.

Progress has not reached as far in the other side of the spectrum.
Stellar evolution models explain naturally the existence of canonical horizontal branch (HB) stars,
but the formation mechanism and the evolutionary path of EHB stars is not well understood. 
EHB stars, being less luminous but hotter than canonical HB stars, are almost invisible in the optical range,
but contribute significantly in the ultraviolet (UV), especially in old stellar populations.
Even though we do not understand well how EHB stars form (for many years these stars were hypothetical), 
the observational fact is that they have been observed in many open clusters \citep{kaluzny92}, globular clusters \citep{catelan09},
and nearby galaxies like M32 \citep{brown00}. See \cite{catelan09} for a review of open problems and the status of observations of EHB stars.

It has long been thought that EHB stars are responsible of the UV upturn phenomenon observed in the spectrum of early type galaxies (ETGs).
This increment in the flux emitted shortward of 2000 {\AA},  also known as UV excess (UVX), was first detected in ETGs by \cite{code79}.
Since then, its origin has been a topic of debate.
\cite{carter11}, \cite{bureau11}, and \cite*{smith12} have shown beyond doubt that the UV upturn is present in passively evolving galaxies with no sign of
recent star formation. 
Thus, the UV upturn must arise from evolved stars, and not from residual star formation.
Were the UV excess produced by young stars, it should show signs of evolution, e.g. the temperature of the UV spectrum should evolve in time.
This is not supported by observations of ETGs \citep{smith12,carter11}.
It is thus fundamental to include properly the evolution of EHB stars in population synthesis models that describe ETGs.

Several mechanisms have been proposed for the origin of EHB stars.
Single star evolution provides one such mechanism.
When a low mass star ascends the red giant branch (RGB), its evolution is governed by the mass of the stellar envelope.
If the mass loss rate is low, the star forms a deep and optically thick envelope, and, when the He flash occurs, the star reaches the HB.
In high metallicity and enhanced He abundance stars, mass loss is more effective and the star may loose all, or nearly all, of its envelope.
The star then becomes bluer and fainter, and reaches the hottest region of the HB, the so called EHB.
RGB stars of all metallicities with envelope mass $< 0.05$ $M_{\odot}$ evolve into the EHB \citep*{dorman93}.
The EHB evolutionary phase is short-lived ($\sim 10^{8}$ yr) \citep{adriano04}.
However, the number of expected EHB stars rises with age.
This suggests that if the UV upturn is related to the presence of EHB stars, the strength of the upturn should increase with metallicity and age \citep{smith12}.

The evolution of binary stars leads naturally to the formation of EHB stars.
\cite{han02,han03} have shown that there are three channels by which interacting binaries may form EHB stars, referred to as hot subdwarf stars in their papers:
{\textit (a)} the stable Roche overflow (RLOF) channel, which results in the formation of hot subdwarf binaries with long orbital periods,
{\textit (b)} the common envelope (CE) ejection channel, which results in the formation of hot subdwarf binaries with short orbital periods, and
{\textit (c)} the merger of two He white dwarfs (2HeWD) to form a single hot  subdwarf star.
In the RLOF channel, the donor star fills its Roche lobe near the tip of its first giant branch ascend, experiences stable mass transfer until its envelope is stripped off, resulting in a naked He core with a thin H envelope.
In the CE ejection channel, the donor star also fills its Roche lobe near the tip of its first giant branch ascend, but dynamically unstable mass transfer leads
to the formation of a CE. The ejection of this CE produces a naked He core with a thin H envelope.
In both cases, the He core ignites to produce a hot subdwarf star.
In the 2HeWD merger channel, a close He WD pair coalesces due to angular momentum loss via gravitational wave radiation.
The merger product ignites He to become a hot subdwarf.  
This binary scenario successfully explains the main observational characteristics of field hot subdwarf stars: 
their distributions in the orbital-period vs. minimun-companion-mass and in the effective-temperature vs. surface-gravity diagrams; 
their distribution of orbital periods and mass function; 
their binary fraction, and the fraction of hot subdwarf binaries with WD companions; their birth rates; and their space density.
In a stellar population, the formation of EHB from binary stars dominates the far-UV part of the population's SED at ages above 1 Gyr (see Figure 9 of \citealt{han07}).
The fraction of EHB stars formed through the 2HeWD merger channel becomes larger than \mbox{50\%} after 10 Gyr (Figure 1 of \citealt{han08}).
This channel produces more massive and more luminous EHB stars, which are the major sources of the FUV flux after 3.5 Gyr (Figure 7 of \citealt{han07}).


Both scenarios for EHB star formation (single and binary) are plausible.
However, the properties of the resulting EHB star depend on the formation channel,
since the distributions of the basic properties of their progenitors are different \citep{han07}.
\cite{smith12} measured the UV upturn as a function of a few spectral indices and stellar velocity dispersion.
They find that the $(F_{UV}-i)$ colour anti-correlates with age, contrary to the predictions of \cite{han07}, and conclude
that binary evolution as the source of the UV upturn is inconsistent with the observations.

In this paper we revisit the inclusion of binary star evolution in population synthesis models,
paying special attention to the predicted UV spectrum.
First, we construct evolutionary tracks for single and binary stars using the \citet{hurley02} public code.
We add to this code the possibility that EHB stars form via the 2HeWD channel proposed by \citet{han02}.
The physical parameters ($T_{\rm{eff}}$, $L$) of the resulting EHB stars are obtained
from the BaSTI database.\footnote[1]{\textit{Bag of Stellar Tracks and Isochrones} (BaSTI)
is a robust and fast interface which uses the evolutionary tracks of \citet{adriano04,adriano06} to obtain
the stellar parameters of HB stars in old stellar populations, including EHB stars.
The tracks by \citet{adriano04,adriano06} are computed with realistic stellar physics. The good agreement between
their theoretical predictions and observations is clear. We do not include their $\alpha$-enhanced stars in our database.}
From our tracks we compute isochrones for stellar populations that include both single and binary
stars, and build the corresponding spectral energy distribution (SED). 
We compare our models with the BC03 models, based on single star evolutionary tracks.

The structure of the paper is as follows. 
In \S\ref{sec:model} we describe how binary star evolution is included in our population synthesis models.
In \S\ref{sec:comparison} we compare the results of this work with previous investigations.
In \S\ref{sec:obs} we analyze recent observations of the old metal rich open cluster NGC 6791 \citep{buzzoni12}, 
contrast them with our model predictions, and explore what this cluster can tell us about the UV upturn in ETGs.
The conclusions of our work are summarized in \S\ref{sec:conclusions}.

\section{A stellar population synthesis model including binary stars}
{\label{sec:model}}

As indicated in \S1, close binary systems may be progenitors of EHB stars through mass transfer or coalescence.
We want to explore the effects of including binary stars, and hence EHB stars, in population synthesis models.
We take a Monte Carlo approach to build from scratch a stellar population formed by both single and binary stars.
To include the binary star population we must specify:
({\textit a}) the mass distribution of the primary (the more massive) and secondary stars in the binary pairs,
({\textit b}) the binary fraction as a function of the mass of the primary star, and
({\textit c}) the distributions of the orbital period and orbit eccentricity of the binary pairs.
For these distributions we follow as close as possible the observations reported in the literature, as explained in
the following subsections.

\subsection{Statistics of binary stars}
\subsubsection{The binary star population}
{\label{sec:parameters}}

The first step in our approach is to generate the mass distribution of the primary and secondary stars in each binary pair.
First, we build the distribution of the mass $M_{1}$ of the primary star.
The values of $M_{1}$ are obtained by populating stochastically the IMF, which we choose, without loss
of generality, to follow the \cite{chabrier03} parametrization:
{\small {
\begin{equation*}
 \xi (\mathrm{log} \, m) \; \propto \;
   \begin{dcases}
     \mathrm{exp}  \, \left[ - \frac{(\mathrm{log} \, m-\mathrm{log} \, m_c)^2}{2\sigma^2} \right] \, ,  & \mathrm{if } \; m \leqslant 1 \, \mathrm{M_{\odot}} \, , \\
     m^{-1.3} \, ,  & \mathrm{if }  \; m > 1 \, \mathrm{M_{\odot}} \, ,
   \end{dcases}
\end{equation*}
}}
\noindent
where  $m_c = 0.08$ M$_\odot$, and  $\sigma$ = 0.69. For the lower and upper mass cutoffs we adopt
\mbox{$m_L = 0.1 \; \mathrm{M_{\odot}}$}, and \mbox{$m_U =100 \; \mathrm{M_{\odot}}$}, respectively.
Each star of mass $M_{1}$ will be the primary star in a pair according to the probability discussed in the
next subsection and listed in Table~\ref{tab:bin_frac}.
Stars in our initial pool that are not selected as binaries remain as single stars.

In order to obtain the mass $M_{2}$ of the secondary star, it is necessary to establish a mass ratio distribution.
\cite{milone12a} studied the properties of photometric binaries in 59 Galactic globular clusters observed with the \textit{HST} WFC/ACS,
and concluded that the distribution of the mass ratio $q = M_2 / M_1$ is almost flat. 
Therefore, we assume that $M_2$ follows from a uniform distribution in $q$:
\begin{equation}
f(q) = 1 \qquad  \mathrm{with} \qquad 0 \leqslant q < 1.
\end{equation}
\noindent
It should be remarked that our procedure does not alter the global IMF.
Adding the binary and single star mass distributions we recover the original IMF. 

\subsubsection{The fraction of binaries}
In the last decades, surveys of binary systems have shown that stellar multiplicity is not the same for all spectral types.
For example, \cite{duquennoy91} find a multiplicity fraction of 0.58 $\pm$ 0.1 for solar type stars, whereas it is 0.42 $\pm$ 0.09 for M dwars \citep{fisher92}.
\cite{lada06} reviewed recent data on stellar multiplicity and concluded that ({\textit a}) most ($\sim$ 69 $\%$) of the stars in the Galaxy are single,
and ({\textit b}) the binary fraction depends on the stellar spectral type of the primary star in such a way that this fraction increases with the
mass of the primary star. Table~\ref{tab:bin_frac} summarizes these results.

\begin{table}
\centering
  \caption{Binary fraction in function of spectral type} \label{tab:bin_frac}
  \begin{tabular}{@{}lllllll}
  \hline
  Spectral & Binary  & Reference \\
  Type      & Fraction &              \\
 \hline
O            &  0.72                  & \cite{mason98}  \\ 
O-B        &  0.65                  & \cite{preibisch99}  \\ 
B-A        & 0.62 $\pm$ 0.2  & \cite{patience02} \\
G-K        & 0.58 $\pm$ 0.1   & \cite{duquennoy91}  \\  
M           & 0.49 $\pm$ 0.09 & \cite{fisher92}  \\ 
Late M    & 0.26 $\pm$ 0.1  & \cite{barsi06} \\
\hline
\end{tabular}
\end{table}

\subsubsection{Orbital parameters}
A binary pair is also characterized by its orbital period and orbit eccentricity.
Surveys of binary systems have shown that the distribution of the orbital period is not uniform.
We assume that the orbital period of our binary pairs follows the gaussian (in log $P$) distribution of period found
by \cite{duquennoy91}:
\begin{equation}
f(\mathrm{log}\, P)=C \mathrm{exp} \left[  - \frac{(\mathrm{log}\, P - \overline{\mathrm{log}\, P})^2}{2\sigma^{2}_{\mathrm{log}\, P}} \right]  ,
\end{equation} 
\noindent
where $P$ is the period in days, $\overline{\mathrm{log}\, P}$ = 4.4, and \mbox{$\sigma_{\mathrm{log}\, P}$ = 2.3}.
For the orbit eccentricity $e$, we assume that it follows a uniform distribution \citep{zhang04, zhang05a}:
\begin{equation}
f(e) = 1 \qquad \mathrm{with} \qquad 0 \leqslant e < 1  .
\end{equation}

\subsection{Evolution of binary stars}
{\label{sec:evbin}}

\setcounter{figure}{0}
\begin{figure*}
\includegraphics[width=0.99\textwidth]{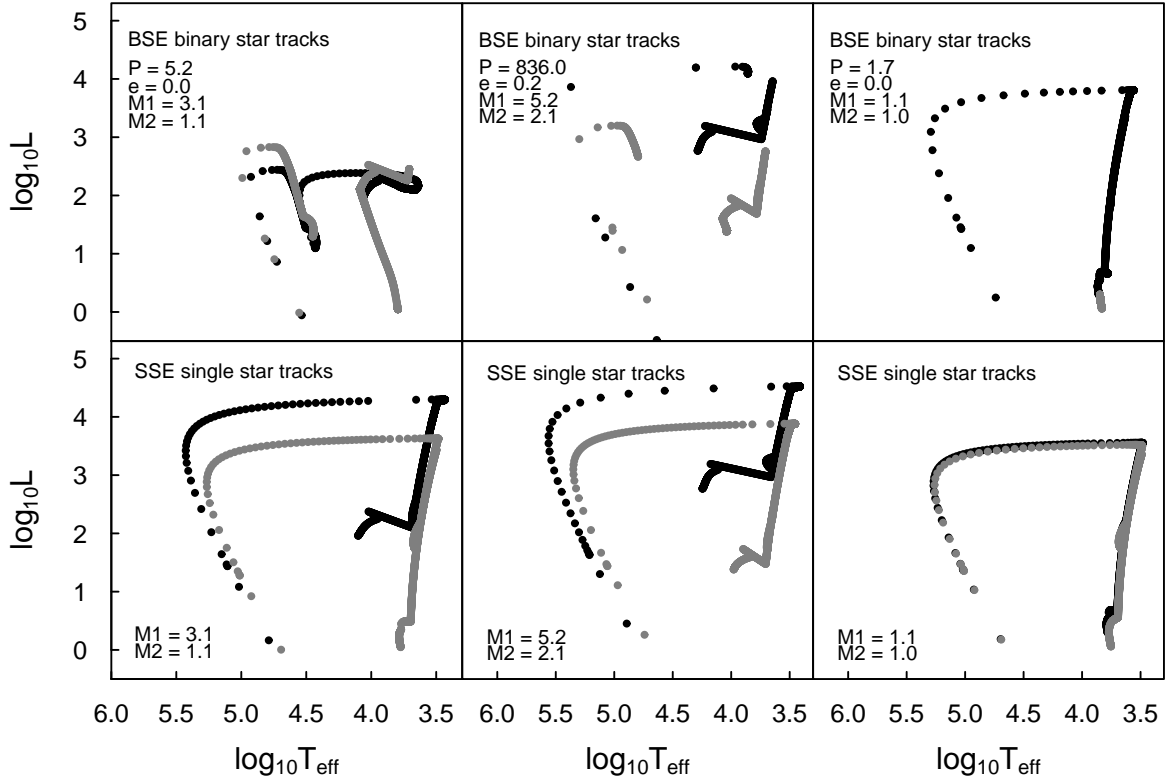}
\caption{Evolutionary tracks computed with the BSE (binary) and SSE (single) stellar evolution code of \citealt{hurley02}.
The top panels show tracks for binary stars with different period ($P$), eccentricity ($e$) and masses. 
Note the differences in the tracks when the same star evolves as a single star (bottom panels). 
The track of the secondary star of the pair is shown in {\textit {gray}} colour.
The stellar masses are given in units of M$_{\odot}$ and the orbital periods are in days.
}
\label{fig:HurleyTracks}
\end{figure*}

\subsubsection{Evolutionary tracks for binary stars}{\label{tracks}}

One important limitation to include binary stars in population synthesis models is the lack
of evolutionary tracks computed {\textit {ex professo}} for binary stars, which are needed
to evolve in time the binary pairs described in \S\ref{sec:parameters}.
In the last decade there has been some important developments in this area, e.g., \cite{hurley02}.
The BSE code developed by \cite{hurley02} provides the evolution of the stellar luminosity, effective temperature, mass, and 
evolutionary phase of both members of a binary pair. The algorithm covers all evolutionary phases from the zero age main
sequence, up to the remnant stage (black hole, neutron star, or white dwarf), and includes the AGB phase.
However, the AGB phase is recognized only by a sharp increase in the star radius following the core He burning phase.
The thermally pulsing ABG stage is not modeled, but its main effects on the long term evolution of the star are taken into account \citep{hurley00}.

The BSE code is designed to model complex binary systems, in addition to model all aspects of single star evolution (SSE).
Important aspects of binary interactions, such as mass transfer, mass accretion, angular momentum loss, supernova kicks,
CE evolution, and the stable RLOF are modeled in the code.
The occurrence of one or more of these processes depends on the parameters describing the binary pair, defined in \S\ref{sec:parameters}:
separation (or period), eccentricity, and mass ratio.
These mechanisms acting in a binary pair, may cause a star of the pair to change its evolutionary path
in the HRD with respect to the path that it would follow as a single star of its initial mass.
The code can compute evolutionary tracks for stars in the mass range from 0.1 to 100 $M_{\odot}$ of metallicity $Z$ from 0.0001 to 0.03.

The orbital period and orbit eccentricity of the pair are input to the BSE code.
In Figure \ref{fig:HurleyTracks} we show an example of how the values of these parameters affect evolutionary tracks. 
In the top left panel of Figure~\ref{fig:HurleyTracks} we show the evolution of an Algol-like system.
The binary interaction leads to the formation of an EHB star (log $L \sim 1-1.5$ and log $T_{\rm{eff}} \sim 4.5$).
At the same time, the secondary star (in grey) increases its mass and becomes a BS, later on evolving to the EHB phase.
It is important to note that the RGB tip is not reached by any of the components of the binary pair.
In contrast, in the bottom left panel (single stars of the same mass as the pair above) both stars reach the RGB tip at log $L \sim 4$.
The single stars do not go through the EHB or BS phase.
The top middle and right panels show two binary systems with different orbital period ($P$), eccentricity ($e$) and mass.
Their single star counterparts are shown in the bottom middle and right panels.
The luminosity at the tip of the RGB is lower if the stars evolve in a binary system.
The origin of these differences is the mass transfer between the stars in binary systems.
A plus of the BSE code is that it can follow the entire evolution of even the most complex binary systems in a small amount of CPU time
(near to one second per system). 
The code is thus ideal to compute tracks for large populations of binaries considering realistically their interactions.

\subsubsection{2HeWD merger}
{\label{sec:merger}}

\cite{hurley02} assume in their BSE code that if 2HeWD stars coalesce, 
the temperature becomes hot enough to start the triple-$\alpha$ process, and the merger transforms into a COWD.
The process of this transformation, i.e., the EHB phase, is not followed in detail.
However, \cite{han02} conclude that under specific conditions the 2HeWD star merger can lead to the 
formation of a single EHB (see also \citealt{webbink84}). 
This mechanism becomes very important in populations older than $\simeq$ 3.5 Gyr (see Figure 3 of \citealt{han10} and Figure 7 of \citealt{han07}).

In order to understand the formation of EHB stars \cite{han02,han03} made a 
detailed investigation of the three main formation channels that can originate an EHB star. 
These channels are: the RLOF, the CE ejection, and the 2HeWD coalescence.

RLOF occurs when a star in a binary pair fills its Roche lobe as a consequence of its evolution, e.g. during the RGB phase, 
or when the system looses angular moment, causing a reduction in the diameter of the orbit.
If mass transfer becomes unstable, e.g., if the donor star is a giant and the companion can not capture all the mass at the 
rate that it is transferred, this mass is accumulated and forms a CE surrounding both stars.
It is then possible that this CE may be lost by the time the two cores spiral in to form a close system. 
The third channel occurs when 2HeWD stars coalesce, and the resulting object ignites He in its core.
This last event is not followed in the \cite{hurley02} BSE code, nor included in their tracks.

To include the 2HeWD star merger in our models, we follow the recipe described by \cite{han02}, \S4, 
which gives the conditions for He ignition in the 2HeWD merger.
The conditions for He to ignite depend on the mass of the more massive component, its accretion history, 
and other parameters related to its thermal structure. 
In order to determine these parameters, \cite{han02} performed a series of calculations that allowed them to follow in detail 
the evolution of the two WDs, the merger, and its products. 
An important result of their work is the mass distribution of the merger products. 
This mass is restricted to a narrow range from $\sim$ 0.4 to $\sim$ 0.65 $\mathrm{M_{\odot}}$. 
This result is not sensitive to the stellar metallicity. 
If a 2HeWD star merger occurs in one of our BSE binary star tracks,
we assume that the mass of the product is the sum of the mass of the two merging WDs at the moment of the coalescence.  
The resulting single star will ignite He if its mass is in the range indicated above.

When He ignition happens in the merger product, we must assign an effective temperature and a luminosity to the star.
This is done by interpolation in mass in a database constructed for this purpose from the output of the BaSTI web tool (ver. 5.0.1).
The resulting effective temperatures are in the range \mbox{20000 K $\lesssim \; T_{\rm{eff}}  \;\lesssim$ 30000 K},
whereas \cite{han02} quote for these stars temperatures in the range of 20000 K to 45000 K.
This difference comes from the fact that in the \cite{adriano04,adriano06} tracks, the range of He core burning masses is narrower than the \cite{han02} distribution. We do not include the lowest and highest values of mass for 2HeWD merger products in  Figure 13 of \cite{han02}. 
For the EHB stars formed through other channels, we use the temperature and luminosity provided by the BSE code.


\subsection{Isochrone synthesis}
{\label{sec:iso}}

\setcounter{figure}{1}
\begin{figure*}
\includegraphics[width=0.99\textwidth]{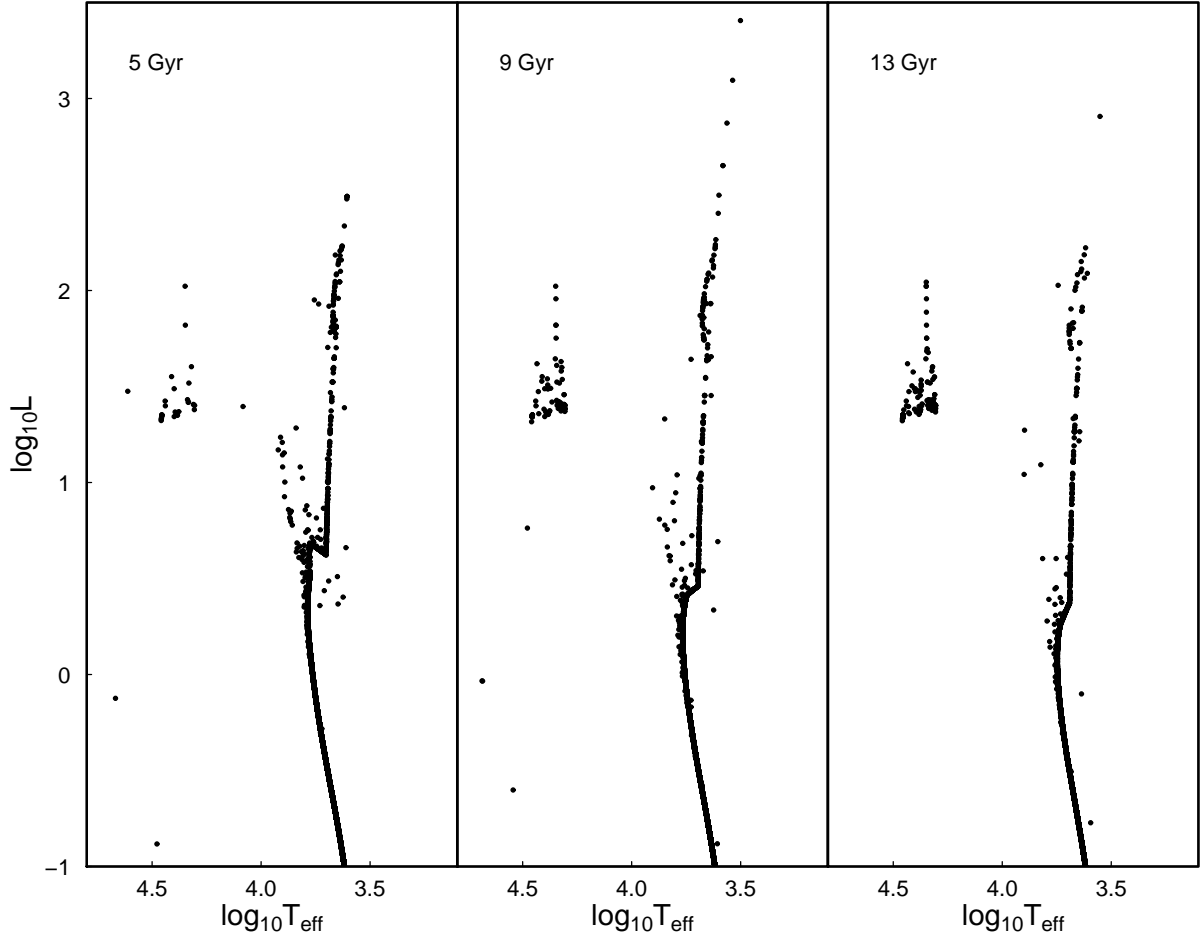}
\caption{5, 9 and 13 Gyr isochrones for solar metallicity derived from the binary and single star evolutionary tracks computed with our implementation of the \citet{hurley02} BSE code, assuming the \citealt{chabrier03} IMF, and the binary fraction described in \S\ref{sec:parameters}. The presence of EHB and BS stars is evident.}
\label{fig:iso}
\end{figure*}
The isochrone describing the position of single and binary stars in the HRD at time \textit{t} is built by interpolation
in our set of BSE evolutionary tracks for the specific metallicity.
Blue Stragglers (BS) are present in these isochrones.
BS are stars which are currently burning H in the core, but their mass is larger than the
turn off mass. They are located above and blueward of the turn off point in a CMD.
This suggests that BS form through a mechanism that permits the star to stay in the main sequence (MS) despite its old age.
There are two possible mechanisms to form BS. Both require that H is replenished in the stellar core by chemical mixing.
One possibility is the coalescence of two MS stars to form a more massive MS star \citep{sig94}.
Other possible mechanism is mass transfer in a binary system \citep{McCrea64}. 
Here we focus on the mass transfer model, included in the BSE tracks. Collisions are important mostly in high stellar density environments \citep{sig94}.
BS have been observed in all stellar systems: globular and open clusters \citep{milone12a,sig94, cenarro10, DeMarchi06}, dwarf spheroidal galaxies \citep{mapelli09}, ultra faint dwarf galaxies \citep{okamoto12}, elliptical and spiral galaxies, and even in our own Milky Way galaxy \citep{monachesi11,clarkson11}. 

Another important aspect of the BSE isochrones is the presence of EHB stars, formed as described above.
It should be noticed that HB stars in single star evolutionary tracks do not reach as hot temperatures as the EHB stars,
even at the lowest metallicities.
In Figure~\ref{fig:iso} we show isochrones corresponding to the evolutionary tracks computed using the recipes described in \S\ref{sec:evbin}.
EHB and BS stars are clearly seen.
The existence of these stars is explained naturally by the evolution of interacting binary stars, which are thought to be present in star clusters and galaxies.
The possibility that EHB stars form via the 2HeWD merger mechanism is still debated, and this hypothesis requires more theoretical and observational support.  
\subsection{Spectral energy distributions}
\label{sec:sed}

\setcounter{figure}{2}
\begin{figure}
\vspace{-8mm}
\begin{center}
\includegraphics[width=0.99\columnwidth]{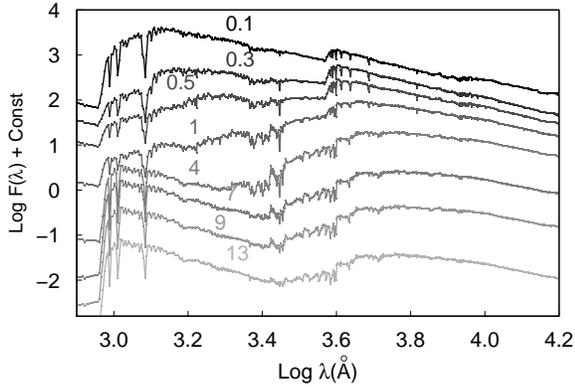}
\caption{Spectral evolution of an SSP of solar metallicity computed with our model. The age in Gyr is indicated next to each SED.}
\label{fig:sed_age}
\end{center}
\end{figure}

The integrated SED of the stellar population is obtained by adding the spectrum of the star at each position in the HRD along the isochrone, 
weighted by the number of stars at this position given by the IMF (see BC03 for details on the isochrone synthesis approach). 
We use the BaSeL 3.1 stellar spectral library \citep{pw02} to compute the spectrophotometric properties of all the models discussed
in this paper. The \citet{chabrier03} IMF is used throughout.

Figure~\ref{fig:sed_age} shows the evolution in time of the SED of a solar metallicity SSP computed from our BSE tracks.
At the youngest ages the UV light is dominated by short-lived massive MS stars, which rapidly evolve and leave the MS, causing a drop in the UV flux.
At ages older than 1Gyr, the evolution of binary systems triggers the formation of EHB stars through the 2HeWD merger mechanism. Even though these
stars are also formed at younger ages through the CE and RLOF channels, its presence is not apparent because massive MS stars are about
100 times more luminous and dominate the UV spectrum. When the massive stars leave the MS, the EHB stars are responsible for the far ultraviolet 
emission. 

The \cite{hurley02} evolutionary tracks show the expected behavior with stellar metallicity, e.g., stellar lifetime increases with metallicity; the evolution
of single low mass stars through the RGB phase determines the position of the star in the HB, at fixed initial stellar mass, lower metallicity stars evolve into
hotter HB stars after the He flash. 
However, the formation of EHB stars through binary interactions depends only on the orbital parameters of the binary pair and not on the stellar metallicity.
Thus, everything else being equal, the number of EHB stars present in a stellar population, should not depend on metallicity. 
This has been reported previously by \cite{han02,han07}. Figure~\ref{fig:spec_z} shows the expected similarity of the SED of SSPs of different
metallicity at 12 Gyr, the UV upturn is clearly seen at all metallicities.

\setcounter{figure}{3}
\begin{figure}
\vspace{-8mm}
\begin{center}
\includegraphics[width=0.99\columnwidth]{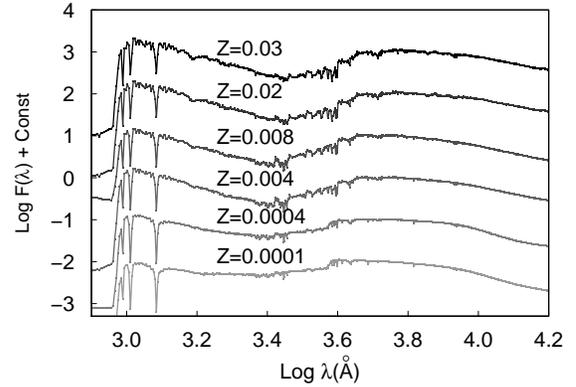}
\caption{SED of SSPs of different metallicities at 12 Gyr. The SEDs have been shifted in the vertical direction for clarity.
The UV upturn short ward of 2000 \AA\ is present at all metallicities.}
\label{fig:spec_z}
\end{center}
\end{figure}

\setcounter{figure}{5}
\begin{figure*}
\begin{center}
\includegraphics[width=0.99\textwidth]{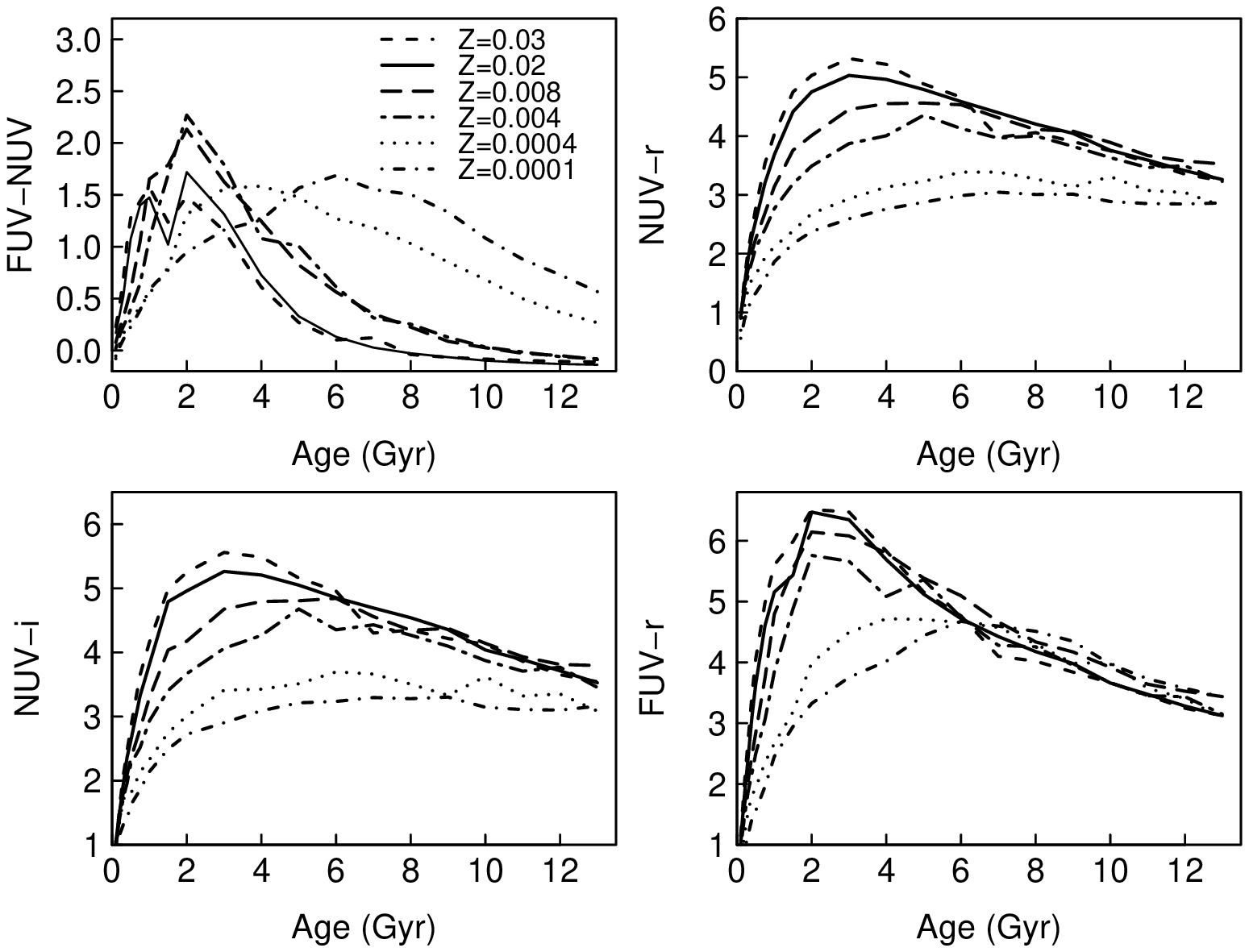}
\caption{UV-optical colour evolution for SSPs of different metallicity. Although the appearance of EHB stars is independent of metallicity,
the presence of an extended HB in the lower metallicity populations increases the NUV flux and makes the FUV-NUV less blue than
for the higher metallicity SSPs.}
\label{fig:colev}
\end{center}
\end{figure*}

In Figure~\ref{fig:colev} we compare the evolution of the UV-optical colours predicted by our models for SSPs of different metallicities.
FUV and NUV refer to the GALEX filters of the same name, and \textit{i} and \textit{r} to the SDSS filters.
The four colours shown in the figure become bluer as the population ages. The behavior of the FUV-NUV colour is quite remarkable.
This colour is dominated at older ages by EHB stars, and becomes very blue for the four higher metallicities shown in the figure.
In the two lower metallicity SSPs, the NUV flux is increased at older ages by HB stars, making FUV-NUV less blue than in their
higher metallictiy counterparts.

\section{Comparison with previous work}
{\label{sec:comparison}}

\setcounter{figure}{4}
\begin{figure}
\vspace{-8mm}
\begin{center}
\includegraphics[width=0.99\columnwidth]{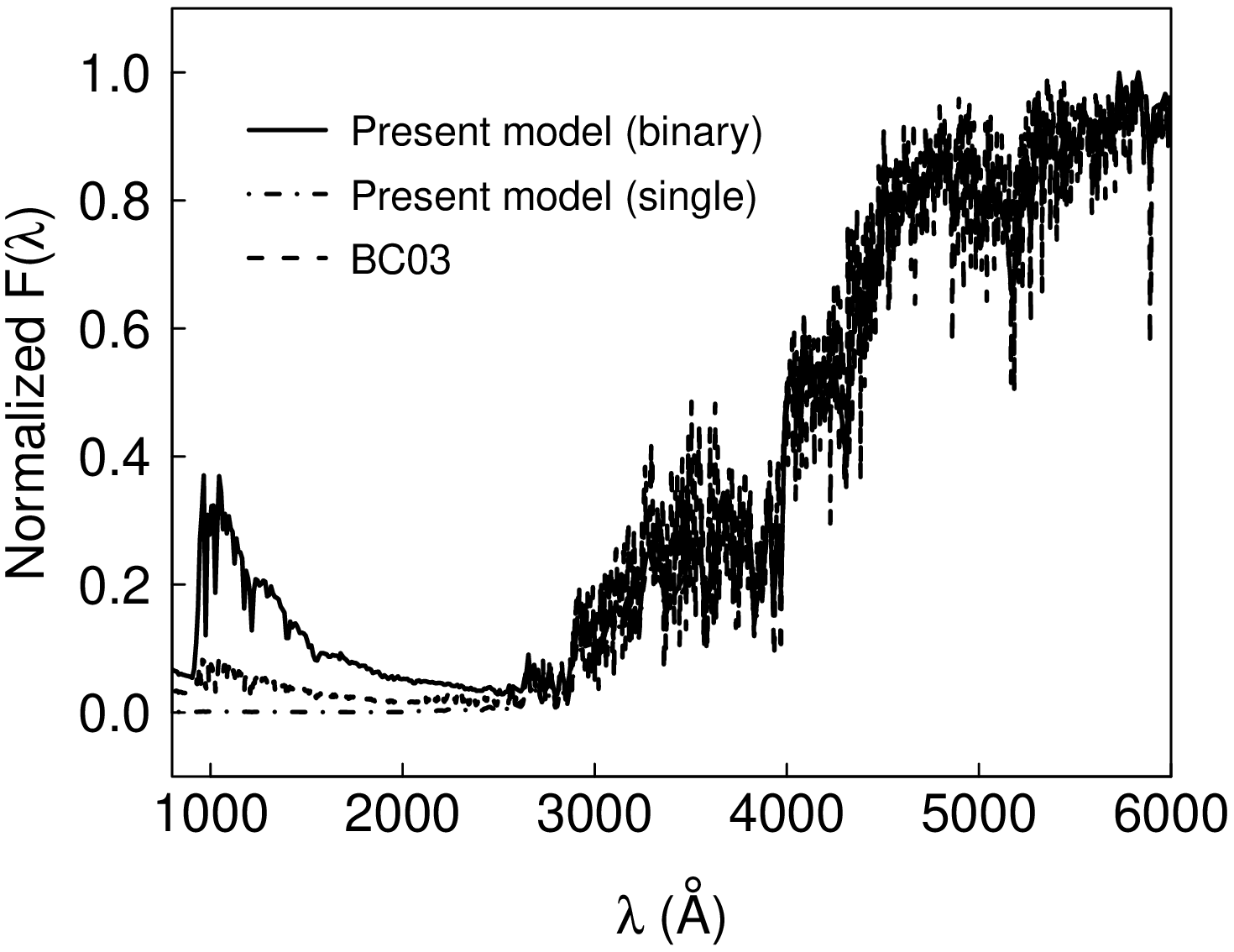}
\caption{
SED of a solar metallicity stellar population at 12 Gyr computed with three different synthesis codes.
From top to bottom in the UV: (\textit{a}) Present models, (\textit{b}) BC03, and (\textit{c}) SED corresponding 
to the evolutionary tracks computed with the \citet{hurley00} SSE code. (\textit{a}) includes and (\textit{b,c}) ignore the evolution of binary stars.
}
\label{fig:spec_comp}
\end{center}
\end{figure}

Most of the population synthesis models available in the literature (\citealt{mf97}, \citealt{cl99}, BC03, \citealt{cm05}) are based on
evolutionary tracks for single stars and do not follow the evolution of the binary stars known to be present in all stellar populations.
Despite this fact, these models reproduce surprisingly well the spectra of most galaxies at all redshifts that have been sampled.
This has been used as an argument in favour of the idea that binary stars are irrelevant in modeling the spectral evolution of galaxies.
However, in the last decade there have been several important efforts to include binary star evolution in population synthesis models.
In this session we compare some results of our models with previous work.

In Figure~\ref{fig:spec_comp} we plot the spectrum of a 12 Gyr old stellar population of solar metallicity computed with three different codes.
It is clear that below $\sim$ 2000 {\AA} the present model shows an upturn in the flux because of the presence of the EHB stars, 
especially compared with the model that uses the SSE tracks, which is completely flat in this wavelength range. 
The BC03 model shows a modest increment in the UV flux because of the contribution of the central star of planetary nebulae (CSPNe) included
in this model, and not in the other two.

\setcounter{figure}{6}
\begin{figure*}
\begin{center}
\includegraphics[width=0.99\textwidth]{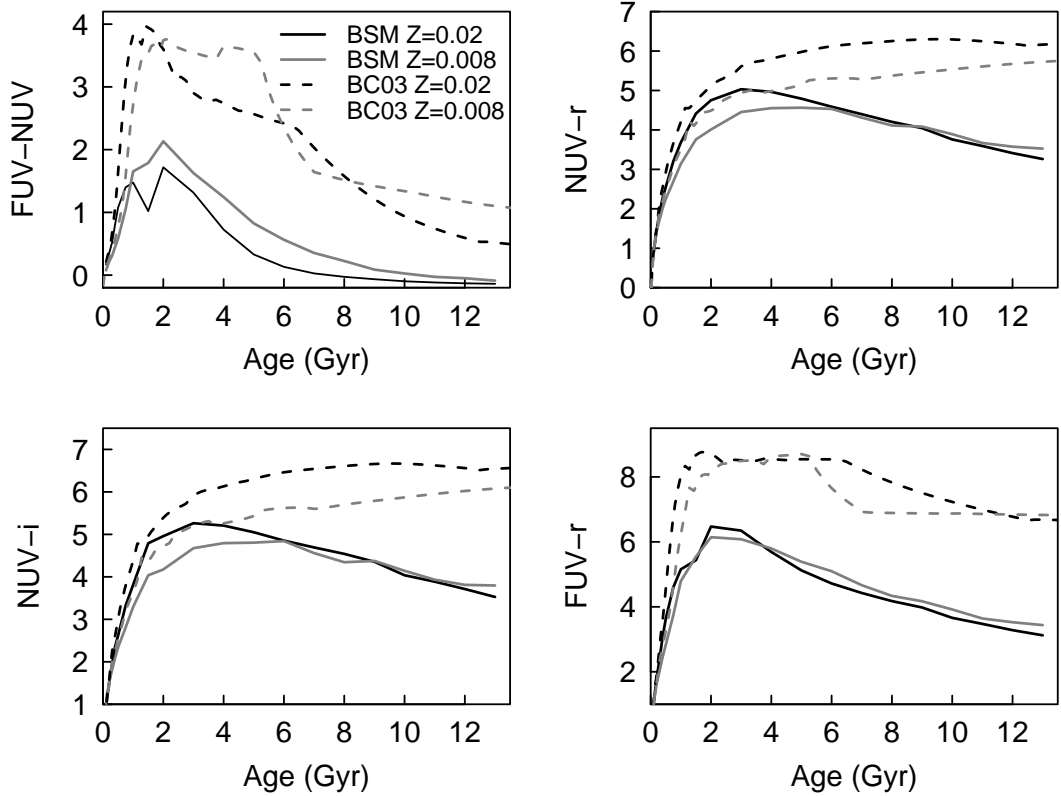}
\caption{Comparison of the UV-optical colour evolution for two of our SSPs with the BC03 models of the same metallicity.
The CSPNe dominate the FUV flux in the BC03 models at late ages and make the model become bluer, but never as
much as our present models, dominated by EHB stars in this wavelength range.}
\label{fig:colev2}
\end{center}
\end{figure*}

Figure~\ref{fig:colev2}, analogous to Figure~\ref{fig:colev}, compares the evolution of the UV-optical colours 
of our models for \mbox{$Z = 0.008$} and 0.02 with the BC03 models for the same values of $Z$. In the NUV-optical colours,
the BC03 models remain red as the population ages, whereas the models with binaries become bluer by more than one magnitude.
In the colours including the FUV filter the BC03 models also get bluer with age, due to the appearance of the CSPNe, but much less than
our present models, dominated in this wavelength region by EHB stars.

\citet{zhang04} and \citet*{zhang05b} also use the \citet{hurley02} evolutionary tracks to follow binary star evolution in their models. 
However, we adopt the default BSE values for the physical parameters governing the RLOF and CE channels.
The CE efficiency parameter ($\alpha_{\rm{CE}}$) is taken as 3.0, the Reimmer mass loss coefficient ($\eta$) is assumed to be 0.5,
and the tidal enhancement parameter is taken as 0.0. \citet{zhang04} and \citet*{zhang05b} use a different set of values. 
The models by \citet{li08} also use the \citet{hurley02} tracks. They assume the binary fraction to be 50\% for all spectral types, and 
examine mostly the optical range and NIR, and compute the Lick indices.
We think that our approach that uses observationally supported distributions of orbital periods and binary fractions makes our models more realistic.

The Binary Population and Spectral Synthesis (BPASS) code \citep{eldridge08, eldridge09, eldridge11} follows the evolution
of fast rotating massive stars in binary systems \citep{eldridge12}, according to their own evolutionary tracks, including the
modeling of the interstellar gas surrounding the stars. Their main goal is to study the effects on stellar lifetime due to the mass loss induced by binary interactions. Our goal is different, since we aim at studying how the integrated properties of stellar populations are modified by the presence of interacting binaries. 
Comparing our models with either the BPASS, the \citet{zhang05b}, or the \citet{li08} models would be an important exercise, since they study the same process
governing different problems.

\section{NGC 6791: an open cluster with UV upturn}
\label{sec:obs}

The open cluster NGC 6791 is an interesting stellar system which shows several particular features: 
(\textit{a}) it lacks RGB stars \citep{kalirai07}, 
(\textit{b}) the mass distribution of its WD cooling sequence is bimodal \citep{kalirai07,bedin08}, and
(\textit{c}) shows bimodality in the HB morphology \citep{buzzoni12}. 
All of these peculiarities are thought to be driven by a mechanism of enhanced mass loss. 
Particularly, \cite{bedin08} have shown that the observed fraction of submassive WDs in this cluster can be naturally accounted for if $\sim$ 34\% of the WDs 
in NGC 6791 are in binary systems.
Additionally, from asteroseismology observations by the \textit{Kepler} space mission, \cite{stello11} found evidence of the existence of unresolved binaries in this cluster.
It is thus plausible that the origin of these peculiarities is related to the evolution of a large number of binary star systems present in the cluster.

The bimodality of the HB morphology in this cluster is relevant for our study. EHB and red clump (RC) stars co-exist in this cluster,
suggesting that a fraction of its RGB stars were subject to enhanced mass loss, evolving into EHB stars, 
while another fraction evolved normally into RC stars. To explore the HB morphology of this cluster,
we compute a synthetic model with the characteristics of this cluster, following the prescriptions of \S\ref{sec:model}. 
The main observational parameters of the cluster are listed in Table~\ref{tab:cluster_param}. The metallicity of
the cluster is Z=0.03.

\setcounter{figure}{8}
\begin{figure*}
\begin{center}
\includegraphics[width=0.99\textwidth]{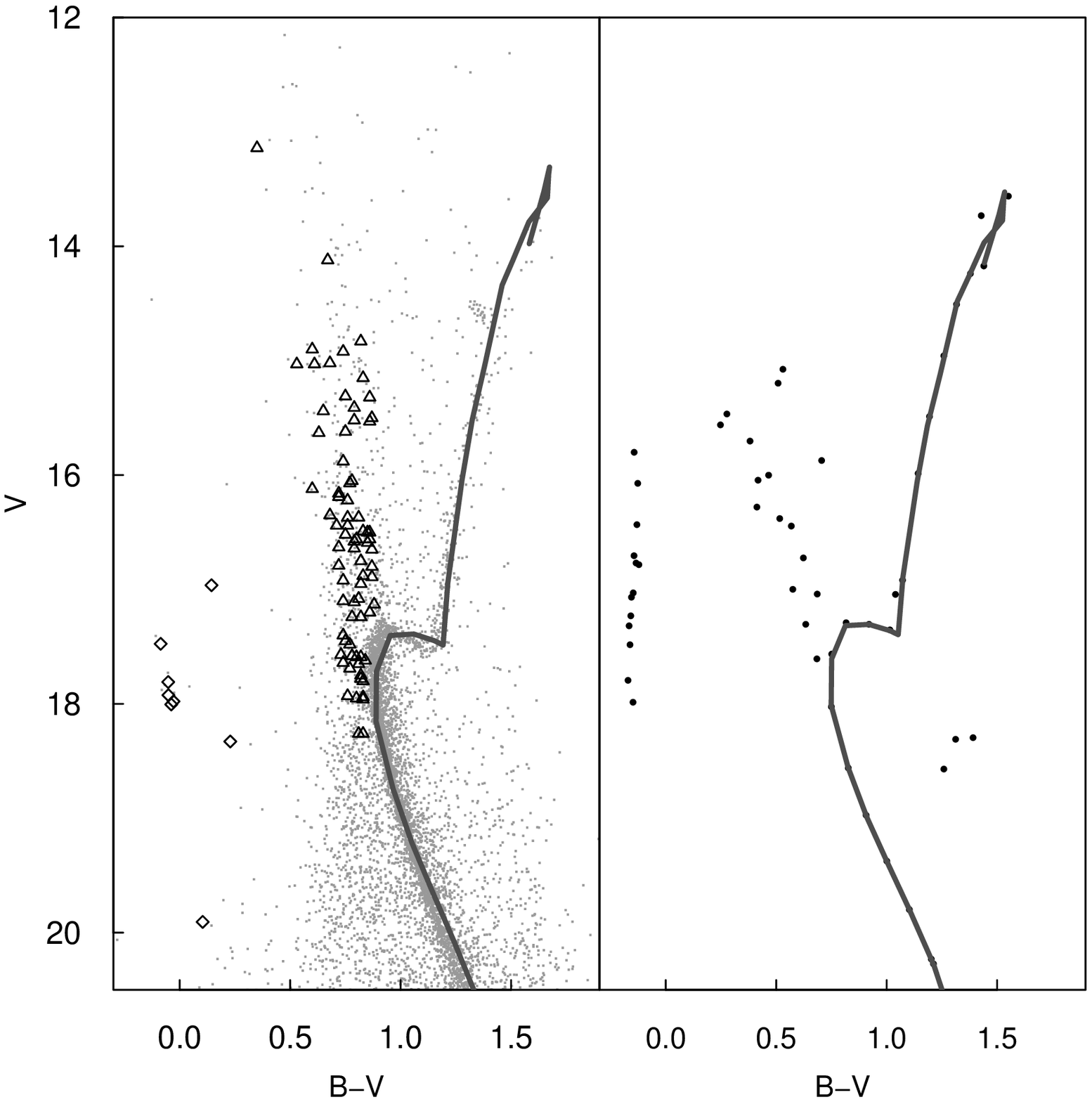}
\caption{\textit{Left panel:} Observed CMD of NGC6791. The small gray dots are the data from the \citet{stetson03} photometric catalog, 
the open triangles are the 75 BS candidates from \citet{al07}, and the open diamonds are the EHB stars from \citet{kaluzny92}.
\textit{Right panel:} Synthetic CMD. The dark gray line (in both panels) shows the isochrone computed for the cluster parameters listed
in Table~\ref{tab:cluster_param}.
}
\label{fig:dcm6791}
\end{center}
\end{figure*}

\setcounter{figure}{7}
\begin{figure}
\vspace{-8mm}
\begin{center}
\includegraphics[width=0.99\columnwidth]{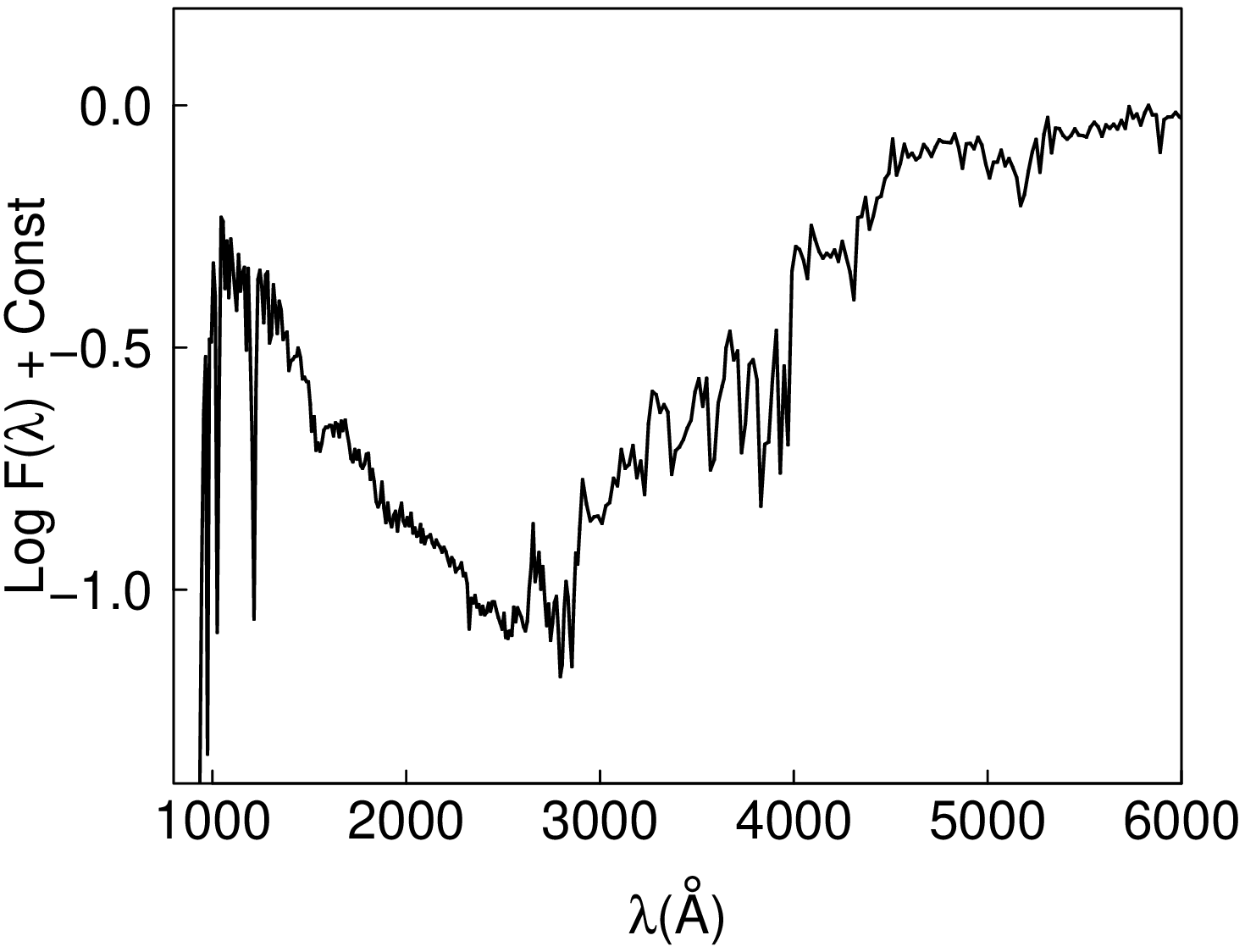}
\caption{SED of NGC 6791 computed with our model.}
\label{fig:sed_ngc6791}
\end{center}
\end{figure}

\begin{table}\centering
  \caption{Open cluster NGC 6791} \label{tab:cluster_param}
  \begin{tabular}{@{}lllllll}
  \hline
  Observational parameter  & Reference \\
 \hline
Age $\sim$ 8.3 Gyr            & \cite{brog12}  \\ 
$\mathrm{[Fe/H]}$ = 0.4 $\pm$ 0.1      & \cite{carraro99,carraro06}  \\ 
$\mathrm{[\alpha/Fe]}$ = solar      & \cite{origlia06}  \\ 
Y = 0.3                       & \cite{brog12} \\
m-M  = 13.51            & \cite{brog11}  \\  
E(B-V) = 0.09-0.18     & \cite{carney05}  \\ 
\hline
\end{tabular}
\end{table}

Figure~\ref{fig:dcm6791} shows that there is very good agreement between the observed (left panel) 
and the synthetic (right panel) CMD of NGC 6791 in $(B,B-V)$.
The position and the extension of the region occupied by the BS are very well accounted for.
The locations of the RC, the RGB and the EHB are also in good agreement in both diagrams.
A remarkable feature is that in our model the bimodality in the HB is evident.
It is important to emphasize that the points located at the right of the MS in the synthetic CMD 
represent stars that are members of a binary system with a BS companion.
These stars have lost a considerable amount of mass, becoming fainter and cooler as they evolve
through the sub giant and RGB phase corresponding to their lower mass.
This suggests that it is possible that the mechanism of enhanced mass loss mentioned above
\citep{kalirai07,bedin08,buzzoni12}, could be driven by binary interactions.
This will explain in a natural manner why only a fraction of the stars passing through the RGB
experience enhanced mass loss, leading to the coexistence of a EHB and a RC normal for this
metallicity and He abundance. 

In our simulated cluster, the fraction of EHB stars with respect to the total of He burning stars is ~ 20\%.
 \cite{kalirai07} argues that $\sim$30\% of the HB stars are hot, which is a extreme value for a high metallicity cluster.
However, \cite{dorman95} using population synthesis models found that for the highest UV upturn galaxies a fraction of 15\% - 20\%
hot HB stars is enough to produce the observed colours. 

\begin{table}\centering
  \caption{Colours of the open star cluster NGC 6791} \label{tab:col_ngc6781}
  \begin{tabular}{@{}lllllll}
  \hline
  Color      &   This model  & \cite{buzzoni12} \\
 \hline
U- B         &   0.602      &  0.60               \\
B - V        &   0.956        &  0.97          \\ 
V - R        &   0.625        &  0.60         \\ 
V - I         &   1.154        &  1.18           \\ 
V - J         &   2.115        &  2.08               \\
FUV - V    &   4.074        &  5.22  \\  
NUV - V   &   4.066        &  5.01    \\ 
\hline
\end{tabular}
\end{table}

\begin{table}
\centering
  \caption{Colours of NGC 6791 vs. ETGs} \label{tab:col_etgs}
  \begin{tabular}{@{}lllllll}
  \hline
                   &   synthetic   &  observed \\
 Color        &   NGC 6791  &  ETGs \\
 \hline
FUV - V       &   4.074                  &  4.95   \\  
NUV - V      &   4.066                   &  4.56     \\
FUV - NUV  &   -0.008                 &  0.4993   \\  
NUV - r       &   4.459                 &   5.3465  \\ 
\hline
\end{tabular}
\end{table}

Figure~\ref{fig:sed_ngc6791} shows the SED corresponding to our synthetic model for NGC 6791, computed as described in \S\ref{sec:sed}.
As expected, the large number of EHB stars present in the simulated cluster contribute to the UV flux, and the spectrum shows
a UV upturn below  \mbox{$\sim$ 2500 $\mathrm{\AA}$}. 
Our spectrum is comparable with the synthetic spectrum shown by \citet{buzzoni12}. 
Their spectrum is built adding the contribution of the stars in the observed CMD of Figure ~\ref{fig:dcm6791}.
The slope and the flux level is similar in both spectra in the UV below \mbox{2000 {\AA}}.
In the optical range, $\lambda \gtrsim$ 4800 {\AA}, our spectrum is somewhat fainter.
This is not surprising because in our model the binary interactions prevent some stars from reaching the RGB tip.
From the synthetic SED we can compute the model cluster colours over the entire wavelength range.
Table~\ref{tab:col_ngc6781} compares the colours computed from our synthetic SED with those listed by \cite{buzzoni12}.
Both sets of colours are in good agreement. 
However, the NUV - V and FUV - V colours are bluer in our model.
The bluer FUV - V colour may indicate a lack of RGB stars in our model.
There are several possibilities to explain a low number of RGB stars.
{\textit i)}
If the donor star in a binary pair is in the sub giant or RGB phase, it may loose enough mass to not be able to reach the tip of the RGB. 
This may be due to the adopted CE ejection efficiency parameter. 
{\textit {ii})}
The initial orbital period distribution or binary fractions adopted in this paper may not be good enough for NGC 6791.
In reality there may be more binary interactions preventing the primary star (star 1) of the pair to reach the RGB.
We should keep in mind that
CE evolution is the most important but the least understood process in binary evolution (\citealt{podsi01}).

The study of old metal rich open clusters may provide an important clue to the study of the UV upturn in ETGs,
since in the clusters it is possible to resolve the hot component of the HB, which could act as a proxy to constrain
the source of the UV excess in ETGs. Recent work by \citet{rosenfield12} shows that emission from EHB stars can
be responsible of the UV upturn in ETGs.
In a subsequent paper \citep{hpb13} we consider a sample of ($\sim$ 3400) ETGs that have been detected in both the
\textit{SDSS}/DR8 and the \textit{GALEX}/GR6 surveys, and analyze the possible scenarios that can influence the
formation of EHB stars in these galaxies, determining the variety observed in the UV spectrum of ETGs.

In Table~\ref{tab:col_etgs} we compare our synthetic cluster colours with the typical UV-optical colour of UV strong ETGs. 
FUV-V and NUV-V are taken from Table 1 of \cite{buzzoni12}, which is constructed from a sample of ETGs from \cite{bureau11} and \cite{buzzoni08}.
The FUV-NUV and \mbox{NUV - r} colours are from \cite{hpb13}.
This sample contains $\sim$ 340 UV strong ETGs with no signs of recent star formation. 
The values listed in the table are the average colours for all of these galaxies.
The synthetic cluster defines a blue limit to the observed averages.

\section{Summary and Conclusions}
{\label{sec:conclusions}}

We have built a population synthesis model that includes, using a simple approach, the evolution of binary stars.
In this model we include the 2HeWD merger channel, suggested by \cite{han02},  for the formation of EHB stars.
An important aspect of our model is the use of the \cite{adriano04,adriano06} evolutionary tracks to estimate
the stellar parameters of the EHB stars produced via the 2HeWD merger process.

The predictions of our model are in good agreement with other well established single star models (e.g., BC03), except
in those instances where the spectrum of the stellar population is dominated by the binary stars or their products (e.g., EHB stars
in the UV of ETGs). 

The CMD diagram predicted by our model for the metal rich open cluster NGC 6791 is in very good agreement with the observed CMD of this cluster.
Our model reproduces well the position and relative number of BS and EHB stars seen in this cluster (Figure~\ref{fig:dcm6791}), as well as its expected SED
(Figure~\ref{fig:sed_ngc6791}) computed by \cite{buzzoni12}.

Despite the good match between model predictions and observations, the formation of EHB stars via the 2HeWD merger process 
is still debated. The frequency of occurrence of this process in real stellar populations remains as an open question.

\section*{Acknowledgments}

We are grateful to the referee, Dr. Zhanwen Han, for valuable suggestions and comments which helped to improve this paper.
Support for this work was provided by the National Autonomous University of M\'exico, through grants IA102311 and IB102212.
FHP acknowledges the hospitality of the UNAM Centro de Radioastronom\'ia y Astrof\'isica during the last stages of this investigation.
FHP acknowledges support from CIDA during her PhD thesis work partially reported in this paper.
This work has made use of BaSTI web tools.

\label{lastpage}
\end{document}